\documentclass[conference]{IEEEtran}
\IEEEoverridecommandlockouts

\usepackage{cite}
\usepackage{amsmath,amssymb,amsfonts}
\usepackage{algorithmic}
\usepackage{graphicx}
\usepackage{textcomp}
\usepackage{xcolor}
\def\BibTeX{{\rm B\kern-.05em{\sc i\kern-.025em b}\kern-.08em
    T\kern-.1667em\lower.7ex\hbox{E}\kern-.125emX}}
\usepackage{comment}
\usepackage{tabularx}
\usepackage{color, soul}
\usepackage{verbatim}
\usepackage{booktabs}
\usepackage{enumitem}

\usepackage{subcaption}
\usepackage{tikz}
\usetikzlibrary{quantikz2}

\usepackage{hyperref}
\hypersetup{
    colorlinks=true,
    linkcolor=blue,
    filecolor=magenta,      
    urlcolor=cyan
    }

\usepackage{booktabs}
\usepackage{multirow}
\usepackage{multicol}

\begin{document}


\title{A Unified Benchmarking Study of Variational Quantum
Eigensolver Ansatz for Electronic Structure Problems}

\author{\IEEEauthorblockN{Anonymous Authors}}

\author{
Julie Maria Raju\textsuperscript{$\Cup$},
Abhoy Kole\textsuperscript{$\Delta$},
Julen Larrucea\textsuperscript{$F$},
Kamalika Datta\textsuperscript{$\Delta$,$\Cup$},
Lennart Weingarten\textsuperscript{$\Cup$},
Rolf Drechsler\textsuperscript{$\Delta$,$\Cup$}\\
\textsuperscript{$F$}\textit{Fraunhofer Institute for Manufacturing Technology and Advanced Materials (IFAM), Bremen, Germany}\\
\textsuperscript{$\Delta$}\textit{German Research Centre for Artificial Intelligence (DFKI), Bremen, Germany} \\
\textsuperscript{$\Cup$}\textit{Institute of Computer Science, University of Bremen, Germany}\\
\{jraju, kdatta, len\_wei, drechsler\}@uni-bremen.de,\\ abhoy.kole@dfki.de, julen.larrucea@ifam.fraunhofer.de
}

\maketitle

\begin{abstract}
Electronic structure calculations become computationally expensive on classical computers as electron correlation increases. Quantum computing offers a promising alternative through hybrid quantum-classical Variational Quantum Eigensolver (VQE) on Noisy-Intermediate Scale Quantum (NISQ) devices. The choice of ansatz plays a pivotal role in  the  performance and scalability of the VQE, yet selecting an optimal ansatz remains challenging. Existing approaches including chemically inspired ansatz, adaptive schemes and Hardware Efficient Ansätze (HEA) lack a unified comparison. This gap makes the scalability and trade-offs of each method unclear for varying molecular sizes. 

This work presents a systematic benchmarking study comparing Unitary Couples Cluster Singles and Doubles Ansatz (UCCSD), fermionic ADAPT-VQE, BasicPCA HEA, and Excitation Preserving HEA across molecules of varying size like $H_2$, $LiH$, $N_2$. We evaluate trade-offs among circuit depth, parameter count, ground-state energy error, and NISQ feasibility. ADAPT-VQE achieves accuracy comparable to UCCSD with lower resource overhead, but becomes impractical for modest system sizes ($\sim$10 electrons) due to increasing circuit complexity. In contrast, HEA reduces hardware costs but requires more parameters and sacrifices accuracy, with performance strongly dependent on molecular size. The Excitation Preserving ansatz remains below chemical accuracy for $H_2$ and $LiH$, and fails for $N_2$, underscoring the need to balance expressibility and trainability. Overall, these results highlight key trade-offs between accuracy and resource efficiency across VQE ansätze, emphasizing the need to improve ADAPT-VQE scalability and HEA accuracy while addressing optimization challenges for near-term quantum hardware.
\end{abstract}


\section{Introduction}\label{sec:introduction}

Over the past decades, extensive research has been devoted to solving the electronic structure problem using both fully quantum algorithms and hybrid quantum-classical approaches. Early work by Whitfield et al.~\cite{Whitfield_2011} employed the \emph{Quantum Phase Estimation} (QPE) algorithm to estimate the ground-state energy of the $H_2$ molecule using the Quantum Fourier Transform. While this approach demonstrated the theoretical potential of quantum computing for chemistry, it relied on precomputed molecular integrals and highlighted significant scalability challenges for practical quantum simulations.

To address these limitations, attention shifted toward hybrid quantum-classical algorithms, particularly the \emph{Variational Quantum Eigensolver} (VQE). Foundational works~\cite{Peruzzo_2014,Yung_2014} established VQE as a promising framework for \emph{Noisy Intermediate-Scale Quantum} (NISQ) devices~\cite{Preskill_2018}. As research progressed, the focus moved toward designing efficient ansatz tailored for electronic structure problems. Romero et al.~\cite{romero2018strategiesquantumcomputingmolecular} provided a comprehensive review of chemically inspired ansatz, demonstrating reduced circuit depth without substantial loss of accuracy. However, subsequent studies~\cite{Grimsley_2019} revealed that the accuracy of trotterized \emph{Unitary Coupled Cluster Singles and Doubles} (UCCSD) can be highly sensitive to operator ordering, particularly in strongly correlated systems.

These limitations motivated the development of alternative ansatz constructions. Extensions such as the \emph{Unitary Coupled Cluster Generalized Singles and Doubles} (UCCGSD) and k-UpCCGSD~\cite{Lee_2018} were proposed to improve accuracy and scalability. In parallel, adaptive approaches such as the \emph{Adaptive Derivative-Assembled Pseudo-Trotter Variational Quantum Eigensolver} (ADAPT-VQE)~\cite{Grimsley_2019_ADAPT} introduced an iterative procedure to construct problem-specific ansatz with a minimal set of operators. Further improvements, including CEO-ADAPT and TETRIS-ADAPT~\cite{Anastasiou_2024,Ram_a_2025}, demonstrated reductions in circuit depth and entangling gates while maintaining chemical accuracy. Nevertheless, recent studies~\cite{hagelueken2026exponentialscalingbarriersvariational} indicate that even optimized adaptive ansatz remain challenging to implement on noisy hardware for larger systems.

This has driven interest in \emph{Hardware-Efficient Ansatz} (HEA), which are designed to align with the native gate structure of quantum hardware~\cite{Kandala2017}. While HEAs significantly reduce circuit depth, they introduce new challenges: as system size increases, the number of variational parameters grows, often leading to optimization issues such as barren plateaus~\cite{Larocca_2025}. Additionally, the absence of physical constraints may result in exploration of unphysical regions of the Hilbert space.

To mitigate these issues, symmetry-preserving ansatz have been proposed to enforce physical constraints during optimization. Such approaches either restrict the circuit to symmetry-conserving subspaces or incorporate penalty terms into the Hamiltonian or cost function~\cite{ryabinkin2018constrainedvariationalquantumeigensolver,NASERI2025115541}. Prior studies~\cite{Lyu_2023,Wang_2024} suggest that directly enforcing symmetries within the circuit is more resource-efficient than penalty-based methods. In this direction, Barkoutsos et al.~\cite{Barkoutsos_2018} introduced particle-number-conserving circuits based on particle-hole representations, utilizing parametrized exchange-type gates~\cite{Roth_2017,Egger_2019,Ganzhorn_2019,Sagastizabal_2019}. This approach enables compact entangler designs with reduced circuit depth and improved optimization stability.
Recent experimental studies on real quantum hardware have reported molecular simulations illustrating both the progress and the remaining hardware limitations of VQE-based chemistry calculations \cite{larrucea2026,belalouiGroundStateEnergyEstimation2025,carrerasLimitationsQuantumHardware2026}.

Despite numerous benchmarking efforts evaluating different aspects of VQE~\cite{Lolur_2021,K_hn_2019,boutakka2025benchmarkingvqeconfigurationsarchitectures,claudino,hu2022benchmarkingvariationalquantumeigensolvers,11378391,10.1063/5.0161057,yin2025benchmarkingimpactactivespace}, most studies focus on specific classes of ansatz in isolation. Consequently, a unified comparison across chemically motivated, adaptive, and hardware-efficient approaches remains limited. Furthermore, the choice of an optimal ansatz for molecular systems of varying sizes is still not well understood.

In this work, we address this gap by conducting a unified benchmarking study across multiple ansatz classes in ideal environment, with particular emphasis on hardware-efficient and symmetry-preserving constructions suitable for NISQ devices. Our analysis spans molecular systems of increasing size, providing insights into the trade-offs between accuracy, circuit complexity, and scalability.

The main contributions of this work are as follows:
\begin{enumerate}
    \item We propose a unified benchmarking framework comparing UCCSD, ADAPT-VQE, and \emph{Hardware-Efficient Ansatz} variants.
    \item We perform a systematic evaluation across molecular systems of increasing size, including $H_2$, LiH, and $N_2$.
    \item We provide a joint analysis of accuracy and hardware cost, considering metrics such as circuit depth, number of two-qubit gates, and variational parameters.
    \item We identify a key research gap between HEA and ADAPT-VQE in balancing accuracy, trainability and NISQ feasibility, highlighting it as a promising direction for developing practical NISQ-compatible quantum simulations.
\end{enumerate}

The remainder of this paper is organized as follows. Section~\ref{sec:background} provides the necessary background on the VQE algorithm and different ansatz constructions. Section~\ref{sec:methodology} describes the methodology, including system setup, ansatz design, and evaluation metrics. Section~\ref{sec:experimental} presents the experimental results and analysis. Finally, Section~\ref{sec:conclusion} concludes the paper and outlines future directions.

\section{Background}\label{sec:background}
In this section, we outline the fundamental principles of the VQE algorithm, particularly the molecular Hamiltonian, the qubit mapping and explores various ansatz designs.

\subsection{VQE Algorithm}

The VQE is a hybrid quantum classical algorithm used to find the minimum eigenvalue of a Hamiltonian $\mathcal{H}$ for an input state $|\psi\rangle$, which is the ground state energy of a molecular system~\cite{Peruzzo_2014}.
\begin{align}
    \mathcal{H}\Psi = E \Psi
\end{align}

According to the Rayleigh-Ritz variational principle, the expectation value of the Hamiltonian for any trial wavefunction provides an upper bound to the true ground-state energy of the system.
\begin{align}
    \langle \psi(\theta)|\mathcal{H}|\psi(\theta)\rangle \geq E_0
\end{align}
where $|\psi(\theta)\rangle$ is a parametrized trial state, $\mathcal{H}$ is the Hamiltonian, $\theta$ denotes the variational parameters and $E_0$ is the ground state energy. The basic workflow of VQE is as follows \cite{McClean_2016}:

\begin{description}
    \item[Classical Preparation:] By providing the molecular geometry the second quantized Hamiltonian can be obtained in the selected basis set.
    \item[Fermions to Qubit Mapping:] Using the qubit encoding the second quantized Hamiltonian is mapped into the qubits.
    \item[Quantum State Preparation:] The ansatz or the trial wavefunction $|\psi(\theta)\rangle$ is prepared by a parametrized quantum circuit $U(\theta)$ with an initial state $|\psi_0\rangle$. Typically the Hartree-Fock state is used as an initial state. There are different ways in which one can construct the ansatz and this will be discussed in detail in later sections.
    \item[Measurement:] The quantum computer runs the state preparation and measurement repeatedly to measure the expectation values of the Hamiltonian.
    \item[Classical Optimization:] Parameters in the quantum circuit is optimized and updated by the optimizer to minimize the expectation value. This cycle continues until convergence.
\end{description}


\subsection{Molecular Hamiltonian}
To find the approximate solution of non relativistic time independent Schrödinger equation, the electronic Hamiltonian for $M$ point charges and $N$ electrons within the Born-Oppenheimer approximation is described by~\cite{Szabo1996}:
\begin{align}\label{first_quantization}
   \mathcal{H}_{elec}= -\sum_{i=1}^{N}\frac{1}{2}\nabla_{i}^{2} -\sum_{i=1}^{N}\sum_{A=1}^{M}\frac{Z_A}{r_{iA}} + \sum_{i=1}^{N}\sum_{j>i}^{N}\frac{1}{r_{ij}}
\end{align}
In the second quantization, this can be expressed using the annihilation $(a_q)$ and creation $(a_p^\dagger)$ operators as:
\begin{align}\label{second_quantization}
    \hat H  
= \sum_{pq} h_{pq}  a_p^\dagger a_q
+\frac{1}{2} \sum_{pqrs} h_{pqrs}  a_p^\dagger a_q^\dagger a_r a_s  
\end{align}

where the indices \textit{p, q, r, s} denote the spin orbitals within the chosen basis set. The terms $h_{pq}$ and $h_{pqrs}$ represents the first and second integral which account for the electronic kinetic energy, electron–nucleus attraction, and electron–electron Coulomb interaction.

\subsection{Qubit Mapping}
In order to execute the second quantized Hamiltonian in the quantum computer the fermionic operators are mapped into the qubits as pauli operators. There are several fermion-to-qubit encoding methods, such as the Jordan–Wigner \cite{jordan1928paulische}, Parity, and Bravyi–Kitaev transformations \cite{jordan1928paulische,seeley_bravyi-kitaev_2012}. This work uses the Jordan–Wigner transformation. For the spin orbital indexed by $p$ the fermionic annihilation and creation operators are mapped as: 

\begin{align}
    a_p &= \bigg (\prod_{k=0}^{p-1}Z_k\bigg)\frac{X_p+iY_p}{2}\\
    a^{\dagger}_p &= \bigg (\prod_{k=0}^{p-1}Z_k\bigg)\frac{X_p-iY_p}{2}
\end{align}

where $X_p$, $Y_p$, $Z_p$ are pauli operators.

There are other strategies that can be implemented to reduce the complexity of the electronic structure problem by reducing the Hamiltonian to a selected subset of orbitals and electrons while treating the remaining core orbitals as frozen and incorporating their effects as an energy shift. 

\subsection{Ansatz Categories}
 Ansatz is the basic architecture of a circuit, a set of gates that act on a specific subsystem. In algorithms like VQE, ansatz is a parametrized trial wavefunction  which is used to approximate the true solution.
There are many ansätze developed to solve the molecular Hamiltonians which can be mainly differentiated into three types.\\

\subsubsection{Problem Inspired Ansatz}
 The problem inspired ansatz usually depends on the specific problem, it can combine the property of the problem and then solve the problem by restricting the search to physically meaningful states.\\
 
 \emph{Unitary Couples Cluster Singles and Doubles} (UCCSD) ansatz \label{UCCSD}: The \emph{Unitary Coupled Cluster} (UCC) ansatz has been widely used to solve the electronic structure calculation using VQE \cite{romero2018strategiesquantumcomputingmolecular}. This is inspired from the classical method~\cite{Taube}. The UCC ansatz can be written as
\begin{align} \label{pauli-exponential}
    |\psi(\theta)\rangle =e^{\hat{T}(\theta)-\hat{T}^\dagger(\theta)}|\psi_0\rangle
\end{align}
where $|\psi_0\rangle$ is the initial state, usually the Hartree-Fock state. The operator $\hat{T}(\theta)$ is the coupled cluster excitation operator which is the sum of excitation at different levels. This is truncated into single and double excitation state for UCCSD ansatz, where 'SD' denotes Singles and Doubles and are represented as:

\begin{align}
    \hat{T}_1(\theta) &= \sum _{i,a}\hat{t}^{a}_i = \sum\theta^a_{i}\hat{a}^\dagger_a\hat{a}_i\\
    \hat{T}_2(\theta) &= \sum_{i,j,a,b} \hat{t}^{ab}_{ij} = \sum\theta^{ab}_{ij}\hat{a}^\dagger_a\hat{a}^\dagger_b\hat{a}_i\hat{a}_j
\end{align}

where $a$ and $a^\dagger$ are annihilation and creation operators, respectively. The ground state energy of the molecular system is obtained by minimizing the expectation value of
the Hamiltonian:

\begin{align}
    E = \min_l \langle HF|e^{-(T-T^\dagger)}H e^{(T-T^\dagger)}| HF\rangle
\end{align}

To implement the UCCSD ansatz on a quantum computer, the fermionic excitation operators are mapped to qubit operators using the Jordan–Wigner, Parity or any other transformation. \\

\subsubsection{ADAPT-VQE} \label{Adapt-VQE}
Even though the UCCSD seems to be the well established choice of ansatz, there are few shortcomings. It fails for some systems \cite{osti_1924347} and become ambiguous with the trotterization \cite{Grimsley_2019}. Hence ADAPT-VQE was proposed to tackle these issues \cite{Grimsley_2019_ADAPT}. It is a dynamical VQE algorithm that constructs the circuit iteratively based on the chosen criteria. The basic outline of the algorithm is as follows \cite{Ram_a_2025}:\\

\begin{enumerate}
    \item This is the regular VQE step where the fermionic hamiltonian is transformed into the qubit representation using any encoding strategies like Jordan-Wigner.
    \item Define the operator pool which must be predefined. This will be a collection of operators which are iteratively chosen for the construction of the ansatz.
    \item Initialize the variational state to an appropriate reference state. Usually the Hartree-Fock state is chosen.
    \item 
    For each operator, $A_k$, in the operator pool, the energy gradient $\frac{\partial E}{\partial\theta_k}$ is calculated with respect to the variational parameter $\theta_k$ at zero. If the norm of the gradient is below the predefined convergence threshold $\epsilon$, the process terminates.
    \item The operator $A_n$ corresponding to the largest magnitude energy gradient $\big|\frac{\partial E}{\partial \theta}\big|$
    is selected and appended to the ansatz as an unitary $e^{\theta_nA_n}$, introducing a new variational parameter $\theta_n$.
    \item The variational parameters are then optimized using the VQE to minimize the expectation value of the Hamiltonian.
    \item Go to step 4.\\
\end{enumerate}

The gradient in step 4 can be calculated using the commutator formula in a quantum computer:
\begin{align*}
\frac{\partial\langle \psi^{(n-1)} | e^{-\theta_k \hat{A}_k} \hat{H} e^{\theta_k \hat{A}_k} | \psi^{(n-1)} \rangle}{\partial\theta_k} \bigg|_{\theta_k=0}\\
= \langle \psi^{(n-1)} | [\hat{H}, \hat{A}_k] | \psi^{(n-1)} \rangle
\end{align*}
Since this selection criterion is based on measuring the commutator for each pool operator, the operator pool plays a significant role in circuit efficiency and the accuracy of the results. There are different variants of this algorithm that use alternative selection or termination criteria \cite{Yordanov2021}, or even different operator pools \cite{Ram_a_2025}.\\

\subsubsection{Hardware-Efficient Ansatz} \label{HEA}
While problem inspired ansatz are constructed based on the symmetries of the electronic structure,
some ansatz architectures are generic and problem-neutral, meaning that they can be used generally for any kind of problems on quantum hardware, this type of ansatz are called hardware efficient ansatz \cite{Kandala2017}.

It is designed to match the hardware capabilities than the physical structure of the problem. Generally, HEA consists of repeated layers of parametrized single qubit rotational gates followed by layers of two-qubit entangling gates, such as CNOT and $XX+YY$ gates, which are explored in this work. The $XX+YY$ interaction corresponds to the unitary operator $e^{-i\theta(X\otimes X + Y\otimes Y)} $, whose matrix representation in the computational basis is given by
\begin{align}
\begin{pmatrix}
    1 & 0 & 0 & 0\\
    0 & cos(\frac{\theta}{2}) & -isin(\frac{\theta}{2})& 0\\
    0 &  -isin(\frac{\theta}{2})&cos(\frac{\theta}{2}) &0 \\
    0 & 0 & 0 & 1 \\
\end{pmatrix}
\end{align}

From theoretical view, these ansatz have trade-offs between expressibility and physical structures. Increasing the layers improve the search space and hence improve the results. At the same time, without the problem inspired ansatz construction it may lead to optimization problems like barren plateau \cite{Holmes}. Apart from the generic HEAs there are also implementation of symmetry-preserving hardware-efficient ansatz that explicitly conserve quantities such as particle number and spin \cite{Gard_2020}\cite{Ayeni}. This work focuses on an ansatz called BasicPCA and excitation-preserving hardware-efficient ansätze.

In this work, we investigate the accuracy vs cost trade-offs of different ansätze across molecular systems of increasing complexity, including $H_2$,$LiH$ and $N_2$. We aim to evaluate their scalability and suitability for NISQ era quantum devices by comparing their performance in terms of achievable accuracy and computational resource requirements in an ideal setting.

\section{Methodology}\label{sec:methodology}
\begin{figure}[h!]
    \centering
    \includegraphics[width=0.8\linewidth]{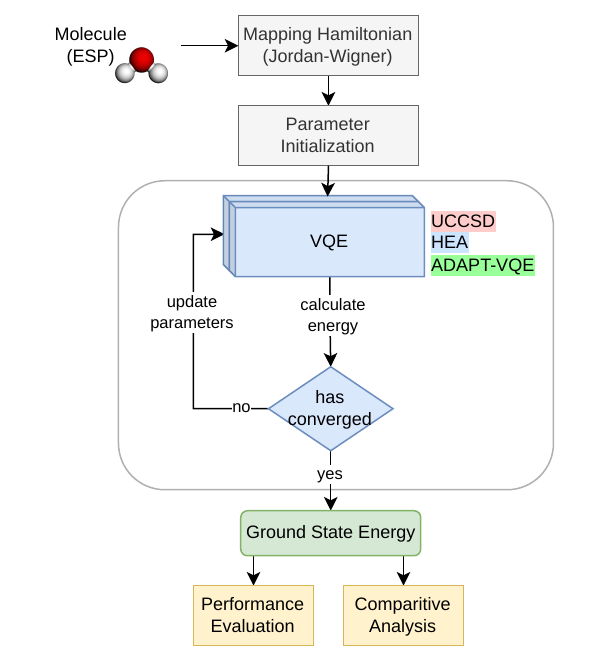}
    \caption{Overall Methodology for Ansatz Analysis}
    \label{fig:methodology}
\end{figure}

In this section, we elaborate the implementation in detail. This includes system setup and Hamiltonian generation, ansatz design, parameter initialization methods, optimization techniques, and performance assessment criteria. The workflow consists of different stages which are illustrated in Figure~\ref{fig:methodology}.

 \begin{figure*}[t] 
    \centering
    \includegraphics[width=0.8\textwidth]{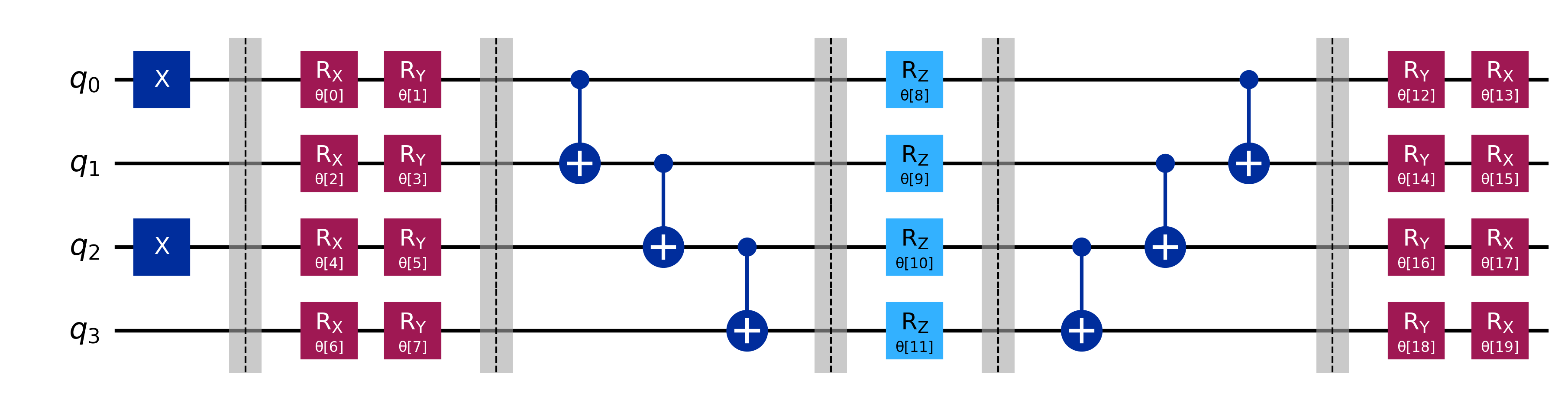}
    \caption{BasicPCA ansatz}
\label{fig:BasicPCA}
\end{figure*}
\begin{figure*}[t] 
    \centering
    \includegraphics[width=0.8\textwidth]{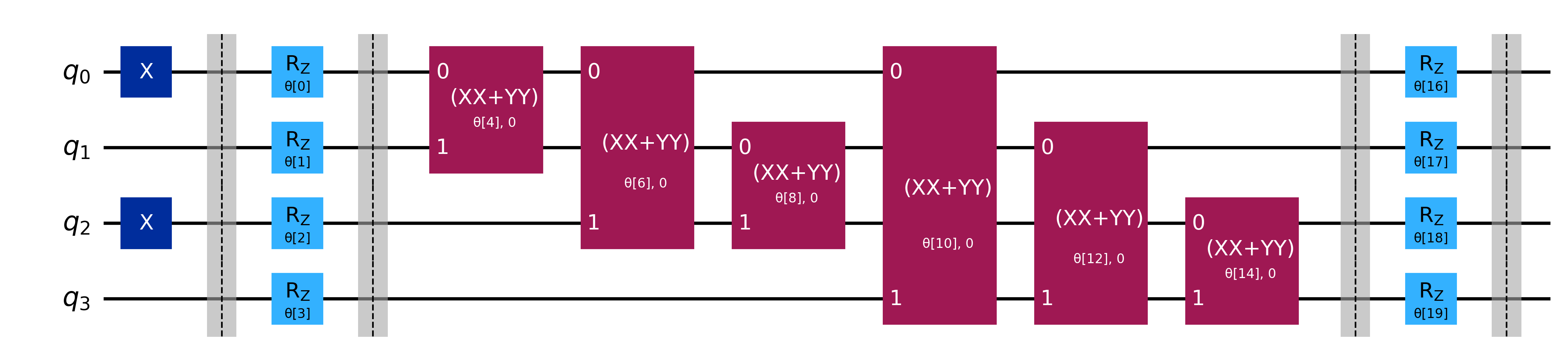}
    \caption{Excitation Preserving ansatz}
    \label{fig:ExcitationPreserving}
\end{figure*}

 \subsection{System Setup and Hamiltonian Construction}
 The VQE is employed to estimate the ground state energy of molecules of varying sizes such as $H_2$, $LiH$, $N_2$ under different variational ansätze.
  For each molecule, the electronic structure problem is formulated using the Born-Oppenheimer approximation using STO-3G basis set. The resulting fermionic Hamiltonian is mapped to qubit Hamiltonian using the Jordan-Wigner transformation which acts as the cost function in the VQE.

 The generated qubit Hamiltonians are further reduced using freeze-core approximation for $LiH$ to make the circuit depth tractable. For consistency the same Hamiltonian is used with all the ansätze considered in this work. 

\subsection{Ansatz Preparation}
In this work, three distinct types of variational ansätze are considered for solving the electronic structure problem. The \emph{Hartree–Fock} (HF) state is used as the reference state for all ansätze considered which is prepared on the quantum circuit by initializing the qubits corresponding to the occupied spin orbitals in the computational basis.

\subsubsection{\textbf{A Problem‑Inspired Ansatz}} Here we use the UCCSD, which is motivated by the standard coupled‑cluster theory in quantum chemistry. The circuit construction is detailed in Section~\ref{UCCSD}. To implement the single and double excitations in the exponential form, such as $e^{-i\theta(P_1 \otimes P_0)}$ as introduced in Eq.~\ref{pauli-exponential}- additional transformations are required to decompose the Pauli gates into the native gate sets. 

    For each qubit on which a Pauli operator other than $Z$ acts, a corresponding basis transformation is applied to rotate the state into the computational $Z$ basis before measurement. For example,
    
        $$ X = HZH , \hspace{0.6cm}
         Y = S^\dagger HZHS $$
    Now the Pauli string is converted into the product of $Z$ operators. Next, a chain of \emph{Controlled-NOT} (CNOT) gates is used to entangle the qubits and hence accumulates the parity of the qubit onto a target qubit. This enables multi-qubit operation to be expressed as a single qubit rotation. Then an $R_z(2\theta)$ is applied to the target qubit, which implements the exponential of the corresponding $Z$ operator. Finally the circuit is uncomputed by reversing the CNOT ladder, followed by the inverse of the initial basis rotations. This allows arbitrary Pauli exponentials to be implemented using a sequence of native single-qubit rotations and two-qubit entangling gates.

\subsubsection{ \textbf{An Adaptive Ansatz}}
In this work, we chose the standard Adapt-VQE approach by using the operator pool composed of single and double excitation operators. These are physically meaningful operators that are mapped into qubit operators using the Jordan-Wigner and is expressed as exponentials of Pauli strings similar to UCCSD but here the algorithm proceeds iteratively as discussed in~\ref{Adapt-VQE}. Here it ensures only operators that significantly reduce the energy below the convergence threshold $\epsilon = 10^{-5}$ are included in the ansatz.

\subsubsection{ \textbf{Hardware‑Efficient Ansätze (HEA)}}
For HEA we considered two variants of ansatz already implemented by Naseri et. al~\cite{NASERI2025115541} and Barkoutsos et. al~ \cite{Barkoutsos_2018}. Both are constructed using hardware efficient gate structures; however, the latter preserves electron number symmetry, while the former does not.

\begin{description}
  \item[BasicPCA:] This ansatz is constructed using a hardware-efficient structure with fixed entanglement blocks and single-qubit rotations. As illustrated in Figure~\ref{fig:BasicPCA}, each layer of the circuit begins with parametrized $R_X$ and $R_Y$ gates applied to every qubit, followed by a forward staircase of CNOT gates. Next, each qubit undergoes individual $R_Z$ rotations, followed by a reverse sequence of CNOT gates, and the layer concludes with parametrized $R_X$ and $R_Y$ rotations. This arrangement creates the necessary entanglement while keeping the circuit depth efficient for near-term hardware.

  \item[Excitation Preserving:] We designed a particle conserving entangler block, as discussed in Section~\ref{HEA} to constrain the wavefunction search to the sector of Hilbert space with a constant number of particles. Each layer starts with single qubit $R_Z$ gates which do not alter the particle number but introduce a phase change which is added to the global phase. Following this, exchange-type $XX+YY$ gates are applied across qubit pairs in a full-entanglement pattern, generating correlations between all qubits while strictly preserving the total number of particles as shown in Figure~\ref{fig:ExcitationPreserving}.
\end{description}

\subsection{Optimization Strategies}

 For the optimization of the VQE algorithm, the $L\_BFGS\_B $ optimizer was employed for all ansatz variants. The initialization strategy for the optimizer varies with the ansatz type. For problem inspired ansatz and adaptive ansatz all initial variational parameters are set to zero while for hardware efficient ansatz multiple initial values were sampled to reduce the risk of trapping in local minima, as highlighted by~\cite{Kandala2017}. Hence in this study the parameters were initialized using a random uniform distribution over the interval ([0, 2$\pi$]). For each bond length, 20 independent optimization runs were performed with different random initializations, and the best result was selected. 
 
 For the dissociation curves, the warm start strategy was employed~\cite{Chai_2025}. The variational parameters for the initial bond length were initialized randomly, and the optimized parameters obtained were subsequently used as the initial values for the following bond lengths.

To evaluate the performance of the ansätze, bond dissociation curves were computed by varying the interatomic distance of the molecule over a predefined range. At each bond length, the ground state energy was obtained using the VQE procedure under identical convergence thresholds and optimization settings. In addition to energy accuracy, a quantitative comparison of circuit resources was performed. For each ansatz, metrics such as the number of variational parameters, number of two-qubit gates, and circuit depth were recorded after convergence. These quantities provide insight into the computational cost and hardware efficiency of the respective approaches.

\section{Results and discussions}\label{sec:experimental}

\begin{table*}[t] 

\begin{tabularx}{1.0\textwidth} { 
  | >{\raggedright\arraybackslash}p{5.3cm} 
  | >{\centering\arraybackslash}X 
  | >{\centering\arraybackslash}X 
  | >{\centering\arraybackslash}X 
  | >{\centering\arraybackslash}X | }
 \hline
 \textbf{Ansatz} & \textbf{\#Qubits} & \textbf{\#Parameters} & \textbf{\#Two-Qubit gates} & \textbf{Circuit depth}\\
 \hline
 \hline
        UCCSD & 4  & 3 & 56 & 112\\
        \hline
         ADAPT-VQE& 4  & 1 & 48 & 96 \\
         \hline
         BasicPCA, rep=1& 4 & 20 & 6 &12 \\
         \hline
         Excitation Preserving, rep=1& 4 & 14 &6  &8 \\
         \hline

\hline
\end{tabularx}
\caption{Comparison of circuit parameters of $H_2$ for different ansätze at equilibrium bond length 0.735\AA.}
\label{table:circuit parameters_h2}
\end{table*}

\begin{table*}[t] 

\begin{tabularx}{1.0\textwidth} { 
  | >{\raggedright\arraybackslash}p{5.3cm} 
  | >{\centering\arraybackslash}X 
  | >{\centering\arraybackslash}X 
  | >{\centering\arraybackslash}X 
  | >{\centering\arraybackslash}X | }
 \hline
 \textbf{Ansatz} & \textbf{\#Qubits} & \textbf{\#Parameters} & \textbf{\#Two-Qubit gates} & \textbf{Circuit depth}\\
 \hline
 \hline
        UCCSD & 10  & 24 & 1616 & 2238\\
        \hline
         ADAPT-VQE& 10  & 10 & 616 & 858 \\
         \hline
         BasicPCA, rep=1& 10 & 50 & 18 &24 \\
         \hline
         BasicPCA, rep=2& 10 & 100 & 36 &47 \\
         \hline
         Excitation Preserving, rep=1& 10 & 65 &45  &20 \\
         \hline
         Excitation Preserving, rep=2& 10  & 120 & 90 & 31\\
         \hline
\end{tabularx}
\caption{Comparison of circuit parameters of $LiH$ for different ansätze at equilibrium bond length 1.595\AA.}
\label{table:circuit parameters_LiH}
\end{table*}

\begin{table*}[t] 

\begin{tabularx}{1.0\textwidth} { 
  | >{\raggedright\arraybackslash}p{5.3cm} 
  | >{\centering\arraybackslash}X 
  | >{\centering\arraybackslash}X 
  | >{\centering\arraybackslash}X 
  | >{\centering\arraybackslash}X | }
 \hline
 \textbf{Ansatz} & \textbf{\#Qubits} & \textbf{\#Parameters} & \textbf{\#Two-Qubit gates} & \textbf{Circuit depth}\\
 \hline
 \hline
        UCCSD & 16  & 315 & 37040 & 47797\\
        \hline
         ADAPT-VQE& 16  & $>$39 & $>$3856 & $>$4875 \\
         \hline
         BasicPCA, rep=1& 16 & 80 & 30 &36 \\
         \hline
         Excitation Preserving, rep=1& 16 & 152 & 120 &32 \\
         \hline

\hline
\end{tabularx}
\caption{Comparison of circuit parameters of $N_2$ for different ansätze at equilibrium bond length 1.1\AA. For ADAPT-VQE, results obtained using an active space transformation (6,6) are reported as a lower bound.}
\label{table:circuit parameters_N2}
\end{table*}

\begin{figure}[t]
\centering
\begin{minipage}[b]{0.48\textwidth}
\includegraphics[width=\textwidth]{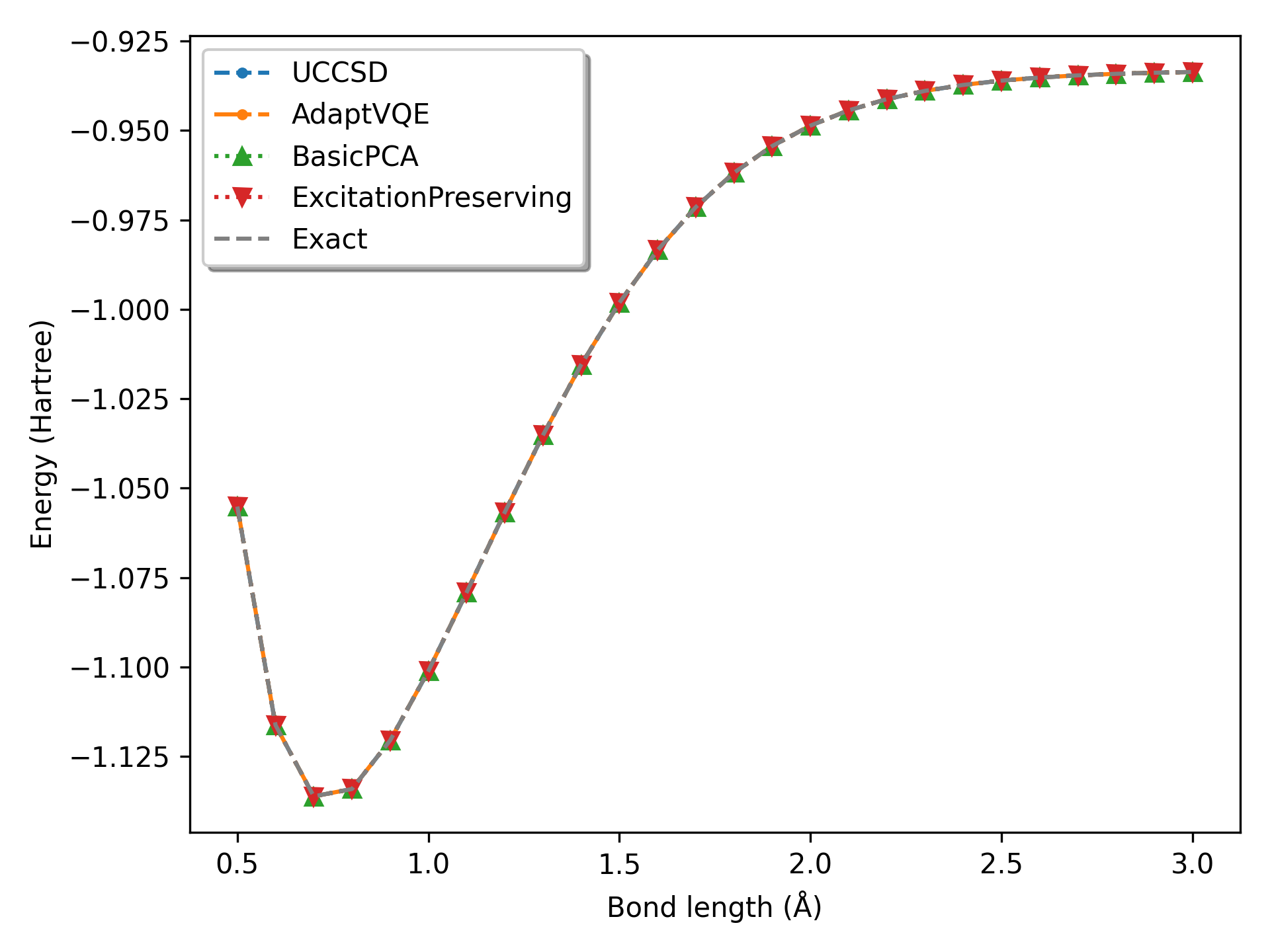}
\caption{Bond dissociation curve of $H_2$}
\label{fig:dissociation_curve_H2}
\end{minipage}
\hfill
\begin{minipage}[b]{0.48\textwidth}
\includegraphics[width=\textwidth]{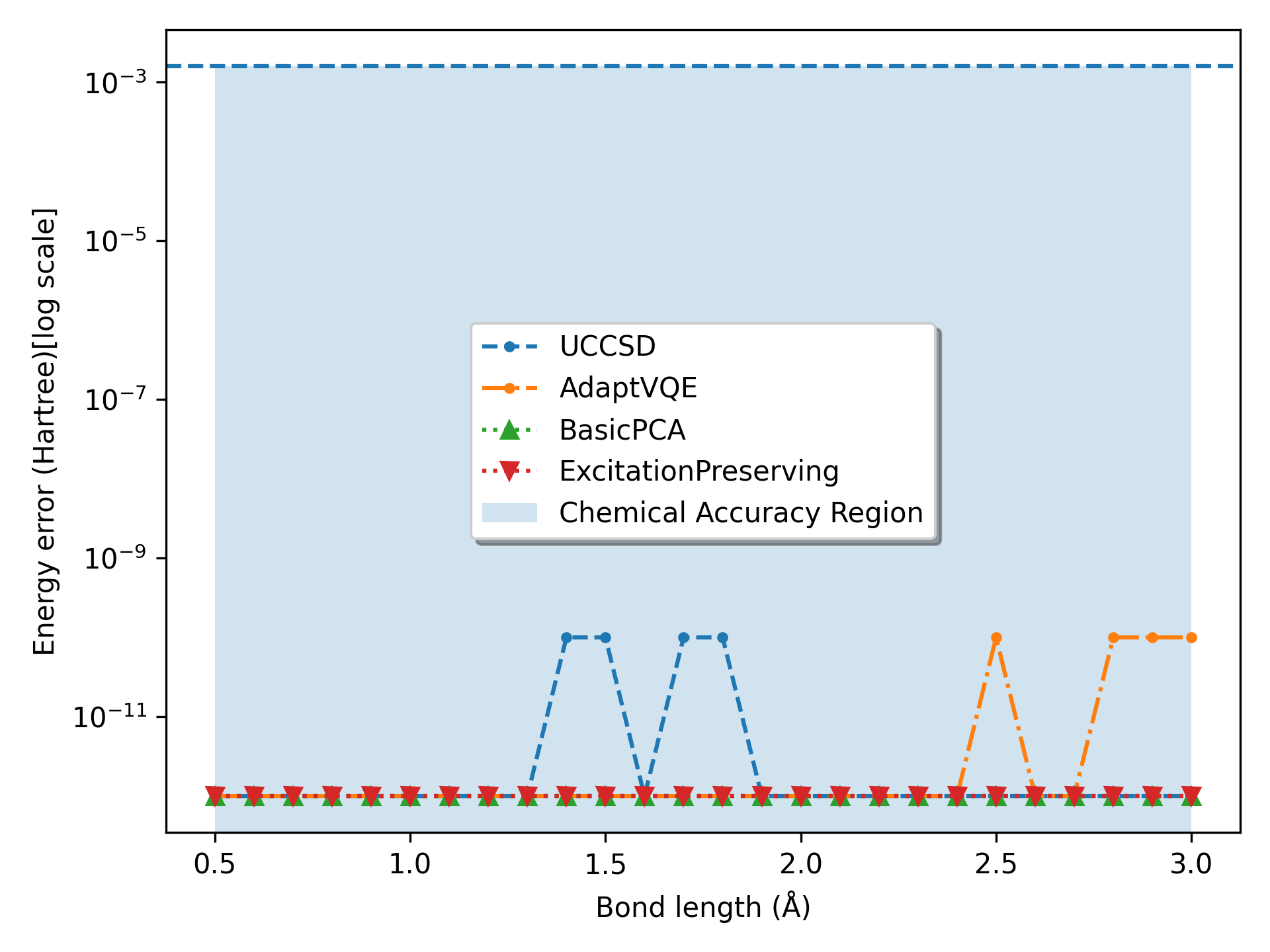}
\caption{Energy deviation from the exact energy of $H_2$}
\label{fig:energy_error_curve_H2}
\end{minipage}
\end{figure}
\begin{figure}[t]
\centering
\begin{minipage}[b]{0.48\textwidth}
\includegraphics[width=\textwidth]{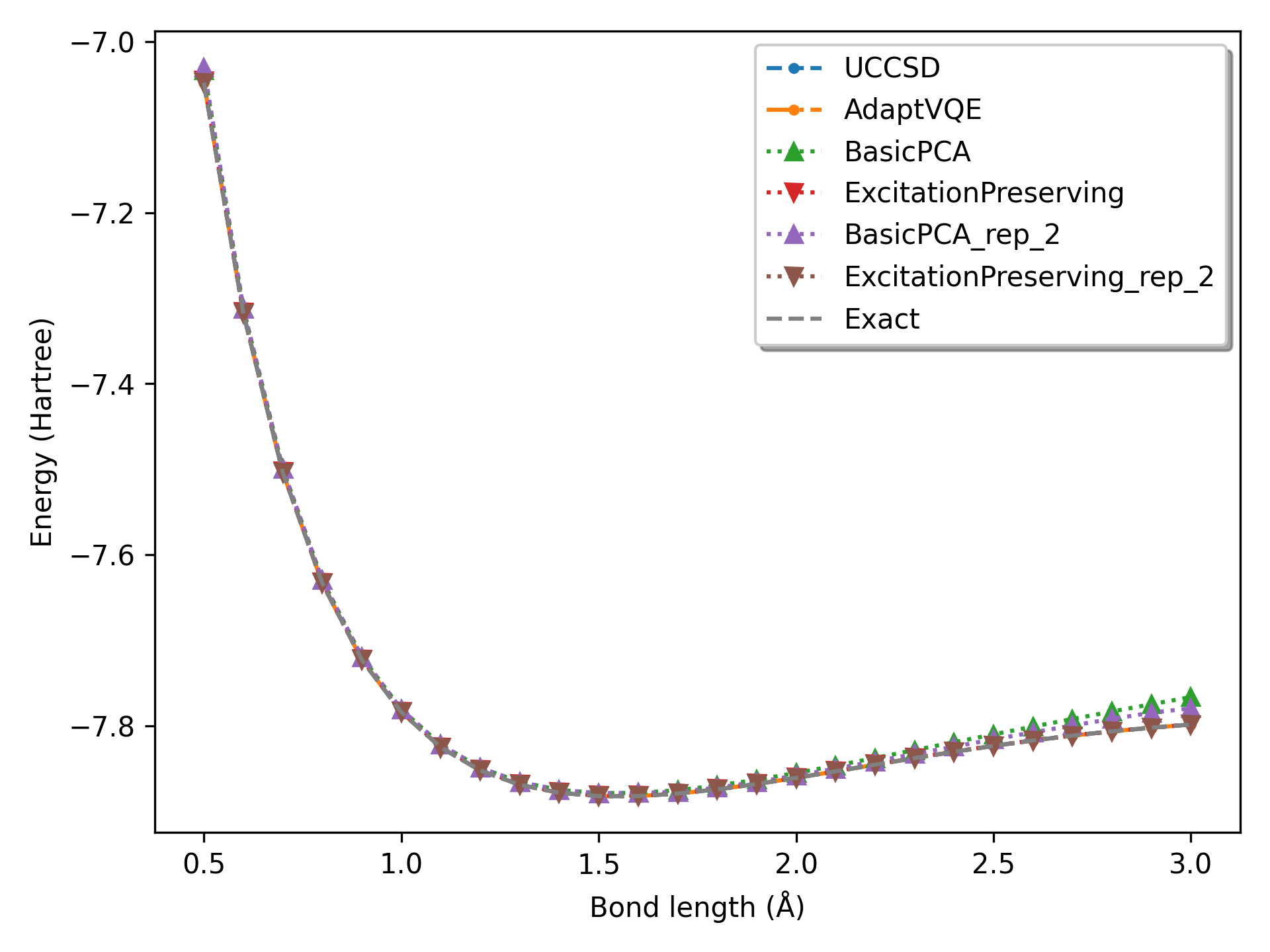}
\caption{Bond dissociation curve of $LiH$}
\label{fig:dissociation_curve_LiH}
\end{minipage}
\hfill
\begin{minipage}[b]{0.48\textwidth}
\includegraphics[width=\textwidth]{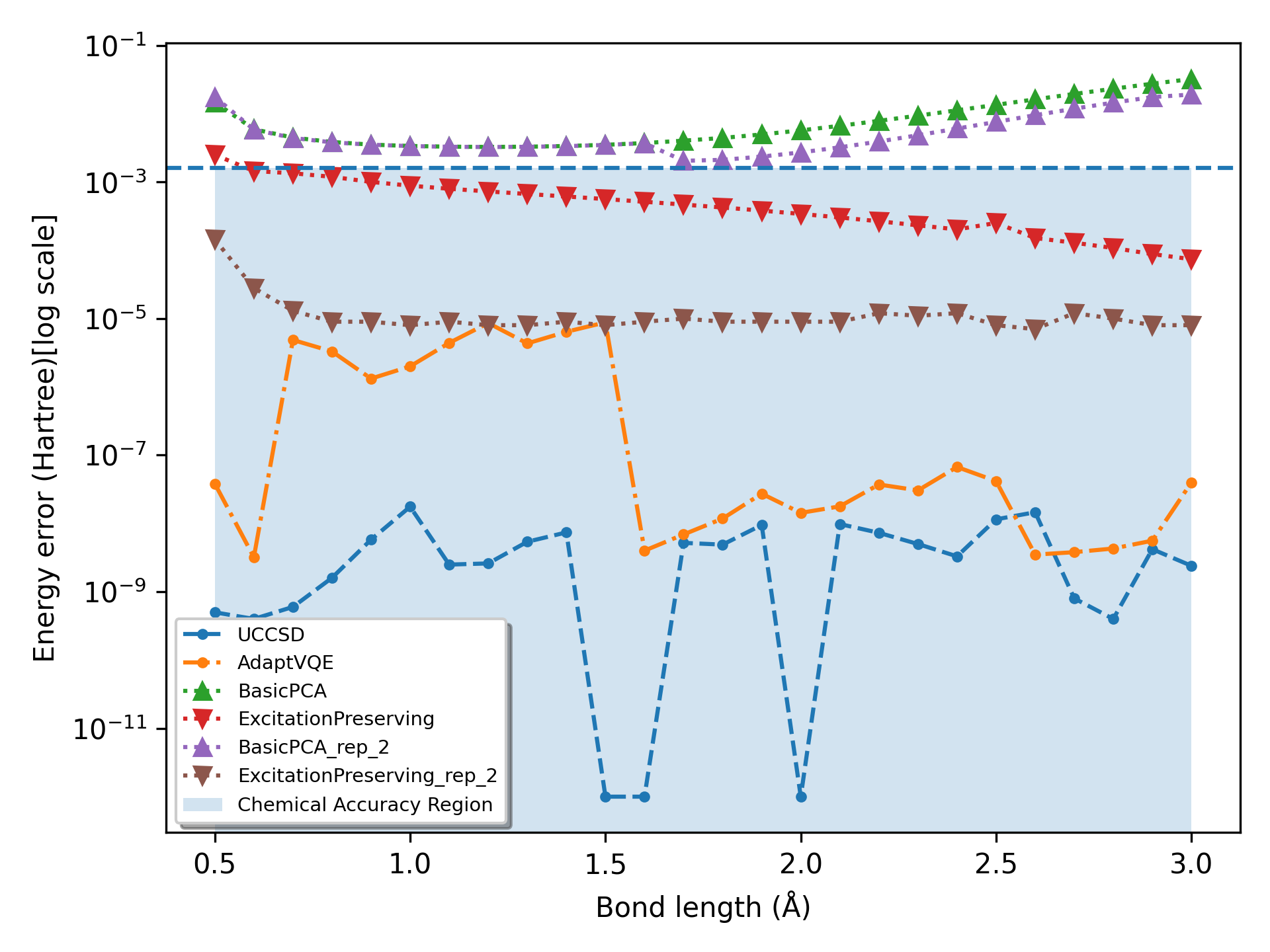}
\caption{Energy deviation from the exact energy of $LiH$}
\label{fig:energy_error_curve_LiH}
\end{minipage}
\end{figure}
\begin{figure}[t]
\centering
\hfill
\begin{minipage}[b]{0.48\textwidth}
\includegraphics[width=\textwidth]{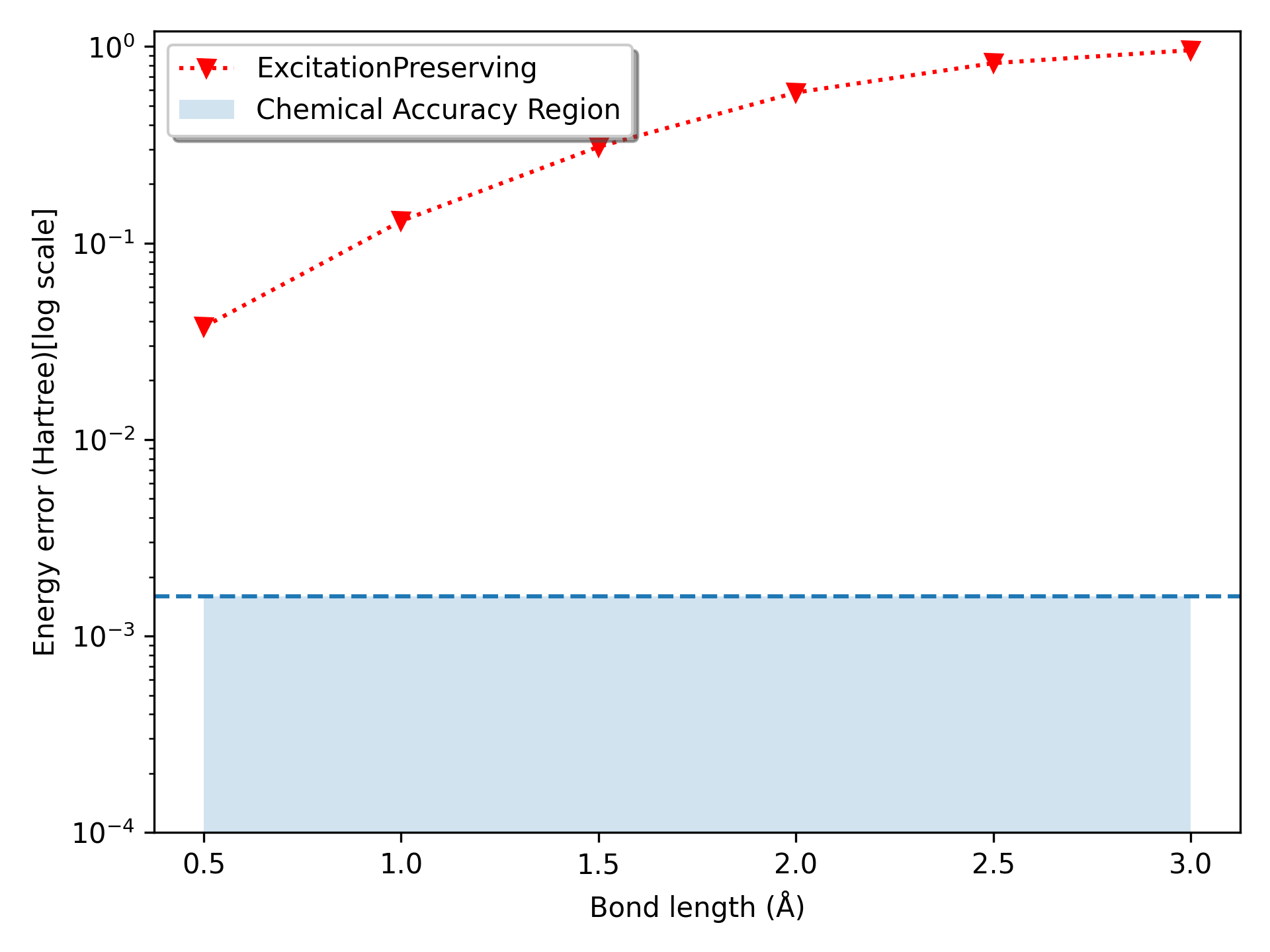}
\caption{Energy deviation from the exact energy of $N_2$}
\label{fig:energy_error_curve_N2}
\end{minipage}
\end{figure}
The results of our experiments comparing different ansätze to find the ground state energy of the molecular systems are summarized and visualized in this section. All the experiments presented are performed on an Intel Core Ultra 7 255U processor with 31 GB main memory, running Ubuntu 24.04.3 LTS on a Windows 10 system.

Quantum circuit construction and simulation were performed using~\texttt{Qiskit} \cite{QiskitCommunity2017}, while molecular geometries were prepared using basis set \texttt{STO-3G} in \texttt{PySCF}~\cite{Sun2017PySCF}. The fermionic Hamiltonians were formulated using \texttt{Qiskit Nature}~\cite{QiskitNature}, and mapped to qubit operators using Jordan-Wigner transformation. All optimizations were carried out using \texttt{L\_BFGS\_B} algorithm on an ideal noise-free quantum simulator. We evaluate the performance of the ansatz on the three molecular systems: $H_2$, $LiH$, and $N_2$.
All results are therefore specific to this methodological setup and should be interpreted within this context, rather than as universal performance comparisons across all VQE configurations.

For $H_2$, we used 4 qubits corresponding to 4 spin orbitals (2 spatial orbitals) in a minimal basis. Since hydrogen has no core electrons, no orbitals were frozen, and the full orbital space was used.

For $LiH$, 10 qubits were used, corresponding to 10 spin orbitals (5 spatial orbitals) with frozen core orbitals to reduce computational cost, as simulations with the UCCSD ansatz became increasingly expensive. The same molecular geometry was maintained across all the ansatz for fair comparison.

For $N_2$, 20 qubits were used, corresponding to 20 spin orbitals (10 spatial orbitals). Even for $LiH$, simulations involving 12 qubits were already computationally demanding. Extending to $N_2$, which requires 20 qubits or 16 qubits with the frozen core approximation becomes impractical, particularly when using the UCCSD or even ADAPT-VQE ansatz. Therefore, for $N_2$, we restricted our analysis to the Excitation Preserving HEA instead.


\subsection{Comparison of Energy Error}
For $H_2$, the ground state energy ranges -1.05 Ha from  -0.93 Ha for bond lengths ranging from 0.5-3.0 \AA. The bond dissociation curve demonstrates how different ansätze perform as the bond is stretched, particularly in the strongly correlated regime. For $H_2$ all ansätze remain indistinguishable as the accuracy remain uniform across all bond lengths as shown in Figure~\ref{fig:dissociation_curve_H2}. To better analyze these deviations, the energy error comparison was studied and is shown in Figure~\ref{fig:energy_error_curve_H2}. Here for each ansatz the absolute value of deviation of energy from exact energy value is calculated for each bond length and is plotted in the log scale to visualize the difference precisely. All ansatz approaches BasicPCA, Excitation Preserving, ADAPT-VQE, and UCCSD achieve very high accuracy for this small molecular system, with energy errors on the order of $10^{-12}$, well below the chemical accuracy threshold, which is conventionally defined as approximately $1\text{kcal/mol}$ ($\approx 1.6 \times 10^{-3}\ \text{Hartree}$).


The ground state energy of $LiH$ ranges from -7.0 Ha to -7.8 Ha for bond lengths ranging from 0.5-3.0\AA.  From Figure~\ref{fig:dissociation_curve_LiH} it is clear that UCCSD, ADAPT-VQE and Excitation Preserving ansatz closely follow the exact energy curve across all bond lengths. In contrast, BasicPCA ansatz exhibits noticeable deviations, which become more pronounced in the strongly correlated region. The resulting error curves show that UCCSD yields the lowest energy error, followed closely by Adapt-VQE as shown in Figure~\ref{fig:energy_error_curve_LiH}. Additionally for hardware-efficient ansätze, circuit depth was scaled by increasing the number of layers. While the Excitation Preserving ansatz achieves chemical accuracy even at low repetitions and improves further with increased depth, the BasicPCA ansatz remains outside chemical accuracy despite increasing the number of repetitions.

For $N_2$, we focus on HEA for bond lengths ranging from 0.5 to 3.0\AA \  with a step size of 0.5, leveraging the fact that UCCSD already provides accurate reference results for this system \cite{xia2020qubitcoupledclustersingles}. As shown in Figure \ref{fig:energy_error_curve_N2}, the Excitation-Preserving ansatz fails to reach chemical accuracy as it exhibits near zero gradients leading to stagnation of the optimization.  This indicates a severe trainability issue. The behavior suggests that the ansatz may not adequately capture the relevant correlated subspace of the $N_2$ triple bonded system, and may be affected by flat regions in the variational landscape.

\subsection{Comparison on Resource Requirements}

 Accuracy alone is not sufficient to determine the optimal ansatz. To assess the scalability and practical feasibility, we further analyzed each ansatz in terms of the number of variational parameters, two-qubit entangling gates, and circuit depth as summarized in Table~\ref{table:circuit parameters_h2}. Excitation Preserving with 14 parameters, only 6 two-qubit and a circuit depth of 8 emerged as the most hardware efficient choice. Similarly, BasicPCA required slightly more parameters (20) but maintained the same two-qubit gate count with a depth of 12, making this as the feasible alternative. In contrast, UCCSD while accurate with few parameters exhibit a circuit depth of 112 and 56 two-qubit gate counts which is roughly 100 times more than the Excitation Preserving ansatz. Such complexity makes it impractical on real hardware. Although ADAPT-VQE is more manageable than UCCSD, it requires more resources compared to HEA.
 
From the perspective of circuit complexity of $LiH$ molecule as shown in Table~\ref{table:circuit parameters_LiH}, UCCSD provides the most accurate results, but incurs significant resource overhead, limiting its applicability on near-term quantum devices.ADAPT-VQE offers a more resource-efficient alternative, yet its requirements remain too large to be fully supported by existing quantum hardware. The BasicPCA ansatz has a relatively low circuit depth and fewer two-qubit gates compared to the Excitation Preserving ansatz. For instance, at one repetition, BasicPCA requires 18 two-qubit gates and a depth of 24, whereas the Excitation Preserving ansatz requires 45 two-qubit gates with a depth of 20. However, this lower resource requirement comes at the cost of reduced accuracy, as BasicPCA fails to reach chemical accuracy even when the number of repetitions is increased. Doubling the repetitions in BasicPCA leads to a proportional increase in gate count and depth (36 gates, depth 47), but does not significantly improve its accuracy.

For $N_2$, Table~\ref{table:circuit parameters_N2} indicates that even with the frozen core approximation (16 qubits), the UCCSD ansatz scales poorly, with a significant increase in circuit depth and two-qubit gate count. While 
HEA offer a more favorable resource scaling, their performance is limited by optimization challenges. In particular, increasing the number of layers to improve accuracy can introduce significant optimization challenges, such as the barren plateau phenomenon, where the simultaneous growth in the number of parameters hinders efficient convergence. For instance, repeating the HEA layer multiple times rapidly increases both the number of variational parameters and two-qubit gates, quickly reaching a scale that is infeasible for current quantum hardware. For ADAPT-VQE, the 16-qubit implementation was not computationally feasible,, since the algorithm requires constructing and evaluating the full circuit after each operator adaptation step, which leads to a prohibitive computational cost. Instead, results were obtained using an active space transformation (6,6) corresponding to a reduced 12-qubit system. This yielded a circuit depth 4875, with more than 3856 two-qubit gates and 39 parameters. Consequently, for the 16-qubit system, the resource requirements are expected to exceed these values significantly.

\subsection{Discussion}
Overall, the results confirm that improved accuracy comes at the cost of increased circuit complexity. Ansätze incorporating enhanced entanglement structures, particularly those inspired by physical principles, demonstrate significant gains in accuracy as circuit depth increases. Furthermore, the choice of entangling patterns and parametrization plays a crucial role in performance, suggesting that the BasicPCA ansatz could achieve improved results with relatively simple structural modifications.

In general, for the molecular systems considered in this study, HEA particularly the Excitation Preserving ansatz demonstrate better scalability while maintaining errors within chemical accuracy . However, the $N_2$ molecule proved to be a notable exception. For this system, the optimization landscape becomes significantly more complex, leading to pronounced trainability challenges that prevent the ansatz from reaching the ground state. 

Although HEA approaches are more scalable and remain feasible for systems in the range of 10 qubits after freezecore reduction, their applicability to larger systems is still limited. As the system size grows, the number of variational parameters increases significantly, leading to more complex optimization landscapes and difficulties such as the onset of barren plateau behavior.

While UCCSD provides a reliable theoretical reference, it becomes computationally prohibitive as system size increases. Even with the application of qubit reduction techniques used in this study, UCCSD becomes impractical for near-term hardware once the requirement exceeds approximately 10 qubits due to the exponential growth in circuit depth and the excessive volume of two-qubit gates.
While ADAPT-VQE offers improvements over UCCSD in terms of resource efficiency as it builds the ansatz adaptively by including only the most significant operators, it remains challenging to implement on near-term devices.This is primarily due to the repeated construction and evaluation of increasingly deep circuits at each iteration, leading to substantial computational overhead and measurement cost, even in light of recent advancements reported in the literature~\cite{hagelueken2026exponentialscalingbarriersvariational}.

Further in this study, we specifically utilized the FreezeCore approximation to make the $LiH$ and $N_2$ simulation feasible within our computational constraints. However, we intentionally refrained from applying additional reduction techniques or symmetry based tapering. This decision was made to preserve the inherent complexity of the problem and to highlight the genuine difficulties current VQE implementations face as system size and correlation strength increase.

\section{Conclusion}\label{sec:conclusion}

In this study, we systematically evaluate UCCSD, ADAPT-VQE, BasicPCA, Excitation Preserving ansätze for estimating the ground state energy of $H_2$, $LiH$, $N_2$ molecules using the VQE with Jordan-Wigner transformation. Our results show that for smaller molecules like $H_2$ all ansätze perform well and considering the scalability, HEA ansätze are the best choice.
For $LiH$, HEA is a pragmatic choice for NISQ-oriented studies where short runtimes and hardware efficiency are prioritized, while UCCSD or ADAPT-VQE is preferred when chemical accuracy targets are dominated, though it may not be feasible on hardware without the active space reduction strategy. For $N_2$,  UCCSD demonstrate clear scalability limitations, supporting these observations. ADAPT-VQE, in contrast, achieves a notable reduction in resource requirements compared to UCCSD by constructing a more compact, problem-tailored ansatz. However, it remains challenging for NISQ devices, as the iterative procedure leads to a growing circuit depth and increased computational and measurement cost with each additional operator. While the HEA is more scalable, it fails to reliably converge to the ground state and does not achieve chemical accuracy in this case. Increasing the expressibility of the HEA improves representational capacity but introduces significant optimization challenges, including barren plateaus and poor convergence behavior, which hinder its ability to reach accurate solutions.

Our results reinforce that tailored ansatz design remains both the primary bottleneck and a central opportunity for advancing quantum computational chemistry. While ADAPT-VQE reduces the resource requirements of UCCSD with minimal loss in accuracy, HEA, despite offering greater scalability, struggles to achieve covergence for larger systems.  Future work will explore larger molecular systems to bridge the gap between current VQE capabilities and the accuracy demands of practical chemical applications. Additionally, efforts will focus on improving ansätze efficiency through strategies such as adaptive physics informed constructions of hardware efficient ansätze or by reducing the number of variational parameters in HEA, which can help mitigate barren plateau issues. While this study was conducted in the idealized noiseless settings, we acknowledge that noise plays a vital role in the practical experiments. Including noise analysis with error mitigation is an intensive study and will be addressed in future work.
\section*{Acknowledgment}
This work was supported in part by the German Federal Ministry of Research, Technology and Space (BMFTR) within the EQeS project under grant no.~01N125050B.

\bibliographystyle{IEEEtran}
\bibliography{bib/books,
bib/reference-abbreviations, bib/reference-esp,
bib/reference-quantum-app, bib/reference-quantum-comp, bib/reference-quantum-decom, bib/reference-quantum-dis, bib/reference-quantum-dyn, bib/reference-quantum-fv, bib/reference-quantum-lmap, bib/reference-quantum-nisq, bib/reference-quantum-opt, bib/reference-quantum-qec, bib/reference-quantum-qem, bib/reference-quantum-tech, bib/reference-larrucea
}

\end{document}